\documentclass[12pt]{article}
\usepackage{a4wide}
\usepackage{amssymb}
\begin{document}
{\renewcommand{\thefootnote}{\fnsymbol{footnote}}
\begin{center}
{\LARGE  Non-bouncing solutions in loop quantum cosmology}\\
\vspace{1.5em}
Martin Bojowald\footnote{e-mail address: {\tt bojowald@gravity.psu.edu}}
\\
\vspace{0.5em}
Institute for Gravitation and the Cosmos,\\
The Pennsylvania State
University,\\
104 Davey Lab, University Park, PA 16802, USA\\
\vspace{1.5em}
\end{center}
}

\setcounter{footnote}{0}

\begin{abstract}
  According to the Belinskii--Khalatnikov--Lifshitz scenario, a collapsing
  universe approaching a spacelike singularity can be approximated by
  homogeneous cosmological dynamics, but only if asymptotically small spatial
  regions are considered. It is shown here that the relevant small-volume
  behavior in solvable models of loop quantum cosmology is crucially different
  from the large-volume behavior exclusively studied so far. While bouncing
  solutions exist and may even be generic within a given quantum
  representation, they are not generic if quantization ambiguities such as
  choices of representations are taken into account. The analysis reveals an
  interesting interplay between ${\rm sl}(2,{\mathbb R})$-representation
  theory and canonical effective theory.
\end{abstract}

\section{Introduction}

Claims such as ``loop quantum cosmology replaces the big bang singularity by a
bounce'' have become commonplace in a large fraction of the literature on the
subject. However, without any qualifications and the specification of
assumptions, they are not supported by current results in this field. The
available evidence is a mixture of numerical and analytical results which are
interpreted as demonstrating the existence of what is referred to as a bounce,
preventing space from approaching a degenerate geometry. However, these
results demonstrate specific features which are realized for a simple
non-singular bounce picture --- a time-dependent volume with a single local
minimum, reached when the energy density is about Planckian --- but do not
necessarily imply it. Quoting this collection of results as proof of
singularity resolution via a generic bounce is therefore misleading.

For instance, numerical solutions for wave functions in simple models may show
a bouncing trajectory of the volume expectation value as a function of
internal time, bounded away from zero by a minimum reached close to the Planck
density \cite{APSII}. But since it is difficult to control a sufficiently
large set of initial states based only on numerics, it is important to know
analytical properties of solutions, at least within certain approximations. On
this side, the main witness of bouncing behavior is usually summoned in the
form of upper bounds on the eigenvalues or expectation values of density
operators \cite{ACS,DensityOp}. Such upper bounds are close to the density at
which numerical solutions reach their minimal volume. They have therefore been
interpreted as generic analytical evidence, strengthening the circumstantial
evidence provided by numerical investigations.

However, the case for a generic bounce constructed so far is incomplete due to
several conceptual gaps in the arguments put forward. For instance, the
existence of upper bounds of the expected densities does not imply that the
expected geometry is never degenerate, if quantum fluctuations are taken into
account: We will present explicit counter-examples in the form of solutions in
which the triad expectation value crosses zero and the density expectation
value always stays below a Planckian upper bound. If the expected triad
vanishes, a fluctuating state is supported on positive and negative triads,
such that the volume (related to the absolute value of the triad variable)
remains positive and follows a bouncing trajectory. However, the geometry may
still become degenerate.  The unquestioned link between bounded densities,
bouncing volume expectation values, and singularity avoidance, made commonly
in arguments in favor of a generic bounce in loop quantum cosmology, therefore
constitues a conceptual gap.\footnote{See for instance the attempted contrast
  between Wheeler--DeWitt quantum cosmology and loop quantum cosmology in
  \cite{ACS}: ``Thus, for a generic state matter density diverges in the
  distant past (or distant future). In this sense the singularity is
  unavoidable in the WDW theory. In LQC by contrast, on a dense subspace the
  expectation value of the volume operator has a nonzero minimum and diverges
  both in the distant past and future. Thus, the density remains finite and
  undergoes a bounce. In this sense the quantum bounce is generic and not tied
  to semiclassical states.'' While the technical comments about loop quantum
  cosmology are correct, they do not rule out a singularity in the sense of a
  degenerate spatial geometry being reached at finite time.}

A further conceptual gap consists in the assumption that only solutions
describing the evolution of a large homogeneous region (a large averaging
volume $V_0$) need be considered because the late-time initial state should
have large-scale homogeneity if it is to approximate our universe. The matter
energy in such a region is large, even if the density is small. In an
effective Friedmann equation, the classical matter term then dominates quantum
fluctuations when the state is evolved toward the big bang, which simplifies
arguments in favor of a bounce. However, as recently pointed out
\cite{Infrared}, the assumption of a large averaging volume fails to describe
the approach to a spacelike singularity in a generic way: While the
Belinskii--Khalatnikov--Lifshitz (BKL) scenario \cite{BKL} does show that
homogeneous models may be used to understand the dynamics close to a spacelike
singularity, such models do not describe an entire homogeneous space but only
an asymptotically small region of decreasing size not bounded from
below. Therefore, small averaging regions, rather than large ones, are
relevant near a spacelike singularity of BKL type. (On approach to a spacelike
singularity, the size of the averaging region must be reduced continually
through infrared renormalization in order to maintain the approximation by a
homogeneous model. A large averaging volume at late times is therefore
consistent with a small averaging volume close to a spacelike singularity.)

For small volumes, it is no longer clear whether the matter term always
dominates quantum fluctuations, and the usual bounce arguments no longer
apply. In fact, it was shown already in \cite{BouncePert} that non-bouncing
solutions may be possible in models of loop quantum cosmology if fluctuations
are sufficiently large. In this paper, we present a more detailed analysis of
the small-volume behavior, using further developments of the methods
introduced in \cite{BouncePert}. In a novel combination with ${\rm
  sl}(2,{\mathbb R})$-representation theory, we will be able to prove that
bouncing solutions are generic within certain representations, including the
one implicitly chosen in \cite{ACS}, but not within the set of all possible
representations.

Models of loop quantum cosmology have been analyzed by a variety of methods,
including algebraic ones and Hilbert-space techniques. While the former are
more general because they are independent of the choice of representations and
can often incorporate quantization ambiguities in a more generic fashion, they
have occasionally been claimed to be less rigorous or even inequivalent to
Hilbert-space results; see for instance \cite{ACS}. Since our new results in
the present paper are based mainly on algebraic methods, we will begin by
demonstrating that the algebraic statements of \cite{BouncePert} are
equivalent to those derived with Hilbert-space techniques, in particular those
used in \cite{ACS}, up to choices of representations and
factor-orderings. This equivalence will serve different purposes in addition
to demonstrating the validity of algebraic results: It will show more clearly
how an analysis of bounce claims is subject to quantization ambiguities, and
it will lead to new analog models in which the bouncing (or non-bouncing)
behavior is shown clearly and intuitively without being obscured by technical
considerations of unobservable features such as the specific form of a
Hilbert-space representation.

By juxtaposing different outcomes for small-volume solutions, including
bouncing and non-bouncing ones depending on properties of quantum states as
well as quantization ambiguities, this paper highlights specific tasks that
remain to be completed before one can claim the robustness or genericness of a
bounce in loop quantum cosmology, if it is in fact realized.

\section{Solvable models}

Loop quantum cosmology \cite{LivRev,ROPP} is a canonical quantization of
homogeneous models of general relativity. Classically, the scale factor $a$
has a canonical momentum given by $p_a=-(3/4\pi G) V_0 a\dot{a}$ where the
time derivative is by proper time and $V_0$ is the coordinate volume of the
homogeneous region chosen to represent all of space. If space is compact,
$V_0$ may but need not be the entire coordinate volume. However, infrared
renormalization implies that $V_0$ must be small compared with the entire
volume if the homogeneous model is supposed to describe the geometry near a
BKL-type singularity.

The canonical variables appear in phase-space functions, such as the Friedmann
equation or Hamiltonian constraint, which may be modified in order to model
quantum effects. When parameterizing quantization ambiguities, it is
convenient to work with a more general set of canonical variables given by
\begin{equation} \label{QP}
 Q=\frac{3(\ell_0a)^{2(1-x)}}{8\pi G(1-x)} \quad,\quad
 P=-\ell_0^{1+2x}a^{2x}\dot{a}\,,
\end{equation}
where $\ell_0$ is such that $\ell_0^3=V_0$ and the parameter $x\not=1$
determines a 1-parameter family of canonical pairs. (In several expressions
derived below the limit $x\to 1$ can be taken and then refers to canonical
variables in which $Q=(3/4\pi G)\log a$.)  The choice $x=-1/2$ is particularly
convenient because it leads to a momentum $P=-\dot{a}/a$ independent of
$\ell_0$, while $Q$ is then proportional to the geometrical volume,
$V=\ell_0^3a^3=4\pi G |Q|$. To facilitate a comparison with \cite{ACS}, we
will often highlight results obtained with this choice of $x$, but note that
it is far from being unique.

We take $Q$ to be a real number, extended from the definition (\ref{QP}) to
negative values by identifying the sign of $Q$ with the orientation of a triad
underlying the spatial geometry determined by $a$. Triads are, in fact,
fundamental geometrical objects in loop quantum cosmology \cite{IsoCosmo}.
A real-valued $Q$, not restricted to be positive, therefore allows for a
one-to-one correspondence with the basic geometrical variable of the theory.

In terms of $Q$ and $P$, the spatially flat Friedmann equation can be written
as
\begin{equation} \label{Friedmann}
 \left(\frac{8\pi G|1-x|}{3}|Q|\right)^{(1+2x)/(x-1)}P^2=\frac{8\pi G}{3} \rho
\end{equation}
with the matter energy density $\rho$. Solvable models in different forms are
obtained for a free massless scalar $\phi$ as the only matter choice, such that
\begin{equation} \label{rho}
 \rho=\frac{1}{2}\frac{p_{\phi}^2}{V^2}= \frac{1}{2} \left(\frac{8\pi
     G|1-x|}{3}|Q|\right)^{3/(x-1)} p_{\phi}^2
\end{equation}
with the canonical momentum $p_{\phi}=\ell_0^3a^3\dot{\phi}$ of $\phi$.
Therefore, imposing the Friedmann equation is equivalent to setting
\begin{equation}
 p_{\phi}^2= \frac{3}{4\pi G} \left(\frac{8\pi G(1-x)}{3}|Q|\right)^2 P^2=
 \frac{16\pi G}{3}(1-x)^2 Q^2P^2
\end{equation}
or
\begin{equation} \label{pphi}
 p_{\phi}= \pm\sqrt{\frac{16\pi G}{3}}\:|1-x| |QP|\,.
\end{equation}

If the scalar $\phi$ is used as internal time, the momentum $p_{\phi}$ plays
the role of a Hamiltonian, generating evolution equations for $Q(\phi)$ and
$P(\phi)$. Up to the absolute value, this Hamiltonian is quadratic according
to (\ref{pphi}), suggesting a coherent quantum behavior. We simplify the
expression for the Hamiltonian by introducing 
\begin{equation} \label{lambda}
 \lambda= \sqrt{16\pi G/3}\:|1-x| \phi
\end{equation}
as internal time, which is canonically conjugate to
\begin{equation}
 p_{\lambda}= \sqrt{\frac{3}{16\pi G(1-x)^2}}\: p_{\phi}
\end{equation}
and therefore implies the Hamiltonian
\begin{equation} \label{H}
 H=p_{\lambda}=\pm|QP|\,.
\end{equation}

\subsection{Holonomy modifications}

Loop quantum cosmology suggests modifications of the
Friedmann equation of two types, inverse-triad corrections \cite{InvScale} and
holonomy modifications \cite{GenericBounce,AmbigConstr}. The former are
motivated by the fact that operator versions of $Q$ in loop quantum cosmology
do not have densely defined inverses, such that there is no direct
quantization of the density (\ref{rho}). Nevertheless, following methods of
the full theory of loop quantum gravity \cite{QSDIII}, it is possible to
construct operators which have an inverse power of $V$ or $Q$ as their
classical limit, but have quantum corrections for small volume. These
inverse-volume corrections imply that the Hamiltonian (\ref{H}) should be
multiplied by a function that approaches one in the classical limit but does
not identically equal one. For our considerations, inverse-volume corrections
will only play a supporting role. Details will therefore be provided in a
later section dedicated to their potential implications.

Models of loop quantum cosmology do not provide an operator version of $P$.
There are, rather, operators that quantize $h_{\delta}=\exp(i\delta P)$ for
any real $\delta$ but are not continuous at $\delta=0$, such that the would-be
operator version of $P=-i{\rm d}h_{\delta}/{\rm d}\delta|_{\delta=0}$ does not
exist. The Friedmann equation, therefore, can be quantized only such that the
classical version is obtained approximately for small $\delta P$, but with
holonomy modifications when $\delta P$ is of the order one. These
modifications are crucial for possible bounces because a simple way of writing
$P$ in terms of $h_{\delta}$ is given by the bounded function
\begin{equation}
 \frac{h_{\delta}-h_{\delta}^*}{2i\delta}= \frac{\sin(\delta P)}{\delta}
\end{equation}
which approaches $P$ for $\delta P\ll 1$ but, unlike $P$, is bounded. For
$x=-1/2$ in (\ref{Friedmann}), the modification implies that the energy
density is always bounded. 

Using this modification, the deparameterized Hamiltonian equals
\begin{equation}
 H_{\delta}=\pm\frac{|Q\sin(\delta P)|}{\delta}
\end{equation}
instead of (\ref{H}). It is no longer quadratic, but still leads to linear
equations of motion for the ${\rm sl}(2,{\mathbb R})$-variables $Q$, ${\rm
  Re}J$ and ${\rm Im}J$ with $J=Qh_{\delta}=Q\exp(i\delta P)$
\cite{BouncePert}. (See also \cite{GroupLQC}. A different ${\rm sl}(2,{\mathbb
  R})$-model of loop quantum cosmology has been investigated in
\cite{CVHComplexifier,CVHPolymer,CVHProtected,CoarseGrainSU11,RenormSU11}.)
The brackets
\begin{equation}
 \{Q,{\rm Re}J\}= -\delta {\rm Im}J\quad,\quad \{Q,{\rm Im}J\}=\delta{\rm
   Re}J\quad,\quad \{{\rm Re}J,{\rm Im}J\}=\delta Q
\end{equation}
are linear, and 
\begin{equation}\label{Hdelta}
 H_{\delta}=\pm\frac{|{\rm Im}J|}{\delta}
\end{equation}
is linear in the generators, up to the absolute value.  The variables are
subject to the condition $Q^2-|J|^2=0$ (implying that $P$ is real) which is
preserved by the evolution equations --- the condition selects a specific
value of the quadratic Casimir of ${\rm sl}(2,{\mathbb R})$.  The model can
be quantized such that the linear ${\rm sl}(2,{\mathbb R})$ relations are
maintained for $\hat{Q}-\frac{1}{2}\hbar\delta$ and $\hat{J}$, ordering
\begin{equation} \label{J}
 \hat{J}=\hat{Q}\hat{h}_{\delta}
\end{equation}
such that
\begin{equation} \label{QJhat}
 [\hat{Q},{\rm Re}\hat{J}]= -i\hbar\delta{\rm Im}\hat{J} \quad,\quad
 [\hat{Q},{\rm Im}\hat{J}]= i\hbar\delta{\rm Re}\hat{J}\quad,\quad [{\rm
   Re}\hat{J},{\rm Im}\hat{J}]=
 i\hbar\delta\left(\hat{Q}-\frac{1}{2}\hbar\delta\right)\,.
\end{equation}
Imposing the quantum reality condition $\hat{Q}^2-\hat{J}\hat{J}^{\dagger}=0$,
such that $\hat{P}$ is self-adjoint, shows that 
\begin{equation}
 \hat{Q}^2-({\rm Re}\hat{J})^2-({\rm Im}\hat{J})^2=
 \frac{1}{2}\left(\hat{Q}^2-\widehat{\exp(-i\delta
     P)}\hat{Q}^2\widehat{\exp(i\delta P)}\right)=
\delta\hbar\left(\hat{Q}-\frac{1}{2}\delta\hbar\right)\,.
\end{equation}
The Casimir operator 
\begin{equation} \label{CasimirJ}
 \left(\hat{Q}-\frac{1}{2}\hbar\delta\right)^2-({\rm Re}\hat{J})^2-({\rm
   Im}\hat{J})^2= -\frac{1}{4}\delta^2\hbar^2
\end{equation}
therefore takes a value which happens to be the limit of the principal
continuous series of ${\rm sl}(2,{\mathbb R})$-representations (contained in
the complementary series as $C_{1/4}^0$ in the notation of
\cite{Bargmann}). The same Casimir, with an operator $\hat{Q}$ having positive
and negative eigenvalues, is obtained for the reducible representation
$D_{1/2}^+\oplus D_{1/2}^-$ where $D_{k}^{\pm}$, are discrete-series
representations with Casimir $k(k-1)$. (An irreducible discrete-series
representation only has $\hat{Q}$-eigenvalues of one sign.) This is in fact
the representation selected if one quantizes $\hat{Q}$ and $\hat{J}$ in
(\ref{J}) via a standard quantization of $\hat{Q}$ and periodic $\hat{P}$.

The quantum Hamiltonian appears in the factor ordering
\begin{equation} \label{Hdelta1}
 \hat{H}_{\delta}= \pm \frac{|{\rm Im}\hat{J}|}{\delta}=
 \pm\frac{1}{2i\delta}\left|\hat{Q}\widehat{\exp(i\delta P)}- 
   \widehat{\exp(-i\delta P)}\hat{Q}\right|\,.
\end{equation}
The availability of two inequivalent representations for the given Casimir has
dynamical implications which will play an important role in our discussions
later on. The spectrum of $\hat{Q}$ is discrete in both cases, but in the
reducible case of $D_{1/2}^+\oplus D_{1/2}^-$ the two subspaces of fixed ${\rm
  sgn}Q$ are left invariant by any ${\rm sl}(2,{\mathbb R})$-element,
including the Hamiltonian. It is therefore impossible for a state supported
only on $Q>0$, say, to evolve into a state with some support on $Q<0$. Such a
representation, unlike the irreducible option of $C_{1/4}^0$, therefore makes
it more likely for initial states to bounce, even though the possibility of
$\langle\hat{Q}\rangle$ approaching zero asymptotically is not ruled
out. However, unless one can show that the reducible representation is somehow
distinguished, ensuring bouncing solutions by a choice of representation would
be ad-hoc. The algebraic treatment is clearly of advantage here because it
highlights possible choices that may be obscured by constructions that start
with the choice of a specific Hilbert space, such as the kinematical one used
in \cite{ACS}. As we will see below, the algebraic approach also allows us to
derive representation-independent statements about solutions.

\subsection{Representations}
\label{s:Rep}

It is straightforward, although ambiguous, to represent the basic operators
$Q$ and $J$ as well as the Hamiltonian $H_{\delta}$ on a Hilbert space. Since
we have already chosen an internal time, such a representation amounts to
deparameterized quantization on a physical Hilbert space. Alternatively, one
may represent $Q$, $J$ (or $h_{\delta}$) as well as $\phi$ and $p_{\phi}$ on a
kinematical Hilbert space and then impose the quantized Friedmann equation via
a constraint. Dirac quantization then leads to the physical Hilbert space. In
\cite{ACS}, group averaging has been applied to complete this procedure,
leading to a representation which at first sight looks rather different from
what one would expect for a quantization of (\ref{Hdelta}). (See also
\cite{LQCScalar} with additional results about non-uniqueness of scalar
products in a Hilbert-space representation.) If $P$ is restricted to the
$2\pi/\delta$-periodicity of (\ref{Hdelta}), $\hat{Q}$ has a discrete spectrum
$\hbar\delta{\mathbb Z}$. The inner product of two wave functions, $\psi_1(Q)$
and $\psi_2(Q)$, on this Hilbert space is given by
\begin{equation}
 (\psi_1,\psi_2)= \sum_{Q\in\hbar\delta{\mathbb Z}} \frac{\psi_1(Q)^*\psi_2(Q)}{|Q|}
\end{equation}
and states obey the evolution equation
\begin{equation}
- \frac{\partial^2 \psi(Q,\lambda)}{\partial\lambda^2}= |\hat{Q}|
\widehat{\frac{\sin(\delta 
     P)}{\delta}} |\hat{Q}| \widehat{\frac{\sin(\delta P)}{\delta}}
 \psi(Q,\lambda) 
\end{equation}
(adapted to our notation and correcting a sign mistake in \cite{ACS}).

Nevertheless, this representation is closely related to a quantization of
(\ref{Hdelta}). First, we can transform to a standard $\ell^2$ inner product
by applying a unitary transformation from $\psi(Q)$ to
$\chi(Q)=\psi(Q)/\sqrt{|Q|}$, such that
\begin{equation}
 (\chi_1,\chi_2)= \sum_{Q\in\hbar\delta{\mathbb Z}} \chi_1(Q)^*\chi_2(Q)\,.
\end{equation}
The evolution equation for $\chi$ is then
\begin{eqnarray}
 -\frac{\partial^2 \chi(Q,\lambda)}{\partial\lambda^2}&=& \sqrt{|\hat{Q}|}
 \widehat{\frac{\sin(\delta P)}{\delta}} |\hat{Q}| \widehat{\frac{\sin(\delta
     P)}{\delta}} \sqrt{|\hat{Q}|} \;\chi(Q,\lambda)\\
 &=& \left(\sqrt{|\hat{Q}|}
 \widehat{\frac{\sin(\delta P)}{\delta}}\sqrt{|\hat{Q}|}\right)^2
\chi(Q,\lambda) \,.
\end{eqnarray}

Solutions of this second-order equation are superpositions of solutions of the
Schr\"odinger-like equation
\begin{equation}
 i\frac{\partial\chi(Q,\lambda)}{\partial\lambda}= \pm \left|\sqrt{|\hat{Q}|}
 \widehat{\frac{\sin(\delta P)}{\delta}}\sqrt{|\hat{Q}|}\right|\chi(Q,\lambda)
\end{equation}
in which the Hamiltonian
\begin{equation} \label{Hdelta2}
 \hat{H}_{\delta}'=\pm \left|\sqrt{|\hat{Q}|}
 \widehat{\frac{\sin(\delta P)}{\delta}}\sqrt{|\hat{Q}|}\right|= \pm
\frac{1}{2i\delta} \left|\sqrt{|\hat{Q}|} \left(\widehat{\exp(i\delta P)}-
    \widehat{\exp(-i\delta P)}\right)\sqrt{|\hat{Q}|}\right|
\end{equation}
is clearly a quantization of (\ref{Hdelta}) in a specific factor ordering
different from (\ref{Hdelta1}). 

In order to see whether this ordering may
imply qualitatively new features, we should relate (\ref{Hdelta1}) to
(\ref{Hdelta2}). Using basic relationships such as
$\hat{h}_{\delta}\sqrt{|\hat{Q}|}=
\sqrt{|\hat{Q}+\hbar\delta|}\hat{h}_{\delta}$, we write
\begin{equation}
 \sqrt{|\hat{Q}|} \widehat{\sin(\delta P)}\sqrt{|\hat{Q}|}= \frac{1}{2i}
 \left(\sqrt{|Q(Q+\hbar\delta)} \:\hat{h}_{\delta}-
 \hat{h}_{\delta}^{\dagger}\sqrt{|\hat{Q}(\hat{Q}+\hbar\delta)|}\right)\,.
\end{equation}
The Hamiltonian (\ref{Hdelta2}) can therefore be written in the form
(\ref{Hdelta1}), but using
\begin{equation}
 \hat{K}=\sqrt{|\hat{Q}(\hat{Q}+\hbar\delta)|} \widehat{\exp(i\delta P)}
\end{equation}
instead of
\begin{equation}
 \hat{J}=\hat{Q} \widehat{\exp(i\delta P)}\,.
\end{equation}
The reality condition 
\begin{equation} \label{RealityJ}
 \hat{J}\hat{J}^{\dagger}=\hat{Q}^2
\end{equation}
is replaced by
\begin{equation} \label{RealityK}
 \hat{K}\hat{K}^{\dagger}= |\hat{Q}(\hat{Q}+\hbar\delta)|\,.
\end{equation}

With this new ordering, the brackets of ${\rm sl}(2,{\mathbb R})$ may be
violated, but only for small $Q$: While the brackets $[\hat{Q},{\rm
  Re}\hat{K}]$ and $[\hat{Q},{\rm Im}\hat{K}]$ are of the correct form, for
the remaining bracket we obtain
\begin{eqnarray}
 [\hat{K},\hat{K}^{\dagger}]&=&\left[\sqrt{|\hat{Q}(\hat{Q}+\hbar\delta)|}
 \widehat{\exp(i\delta P)}, \widehat{\exp(-i\delta P)}
 \sqrt{|\hat{Q}(\hat{Q}+\hbar\delta)|} \right]\nonumber\\
 &=& |\hat{Q}(\hat{Q}+\hbar\delta)|- \widehat{\exp(-i\delta P)}
 |\hat{Q}(\hat{Q}+\hbar\delta)|\widehat{\exp(i\delta P)}\nonumber\\
&=& |\hat{Q}(\hat{Q}+\hbar\delta)|- |(\hat{Q}-\hbar\delta)\hat{Q}|=
\left\{\begin{array}{cl} 2\hbar\delta\hat{Q} & \mbox{if }|Q|\geq\hbar\delta\\
    2\hat{Q}^2{\rm sgn}\hat{Q} & \mbox{if
    }|Q|<\hbar\delta \end{array}\right. \label{KK} 
\end{eqnarray}
where the inequalities for $|Q|$ correspond to the support of a state in the
$Q$-representation on which the commutator acts. 

The new commutation relation (\ref{KK}) together with (\ref{RealityK}) shows
that
\begin{eqnarray}
 |\hat{Q}(\hat{Q}+\hbar\delta)|- ({\rm Re}\hat{K})^2-({\rm Im}\hat{K})^2&=&
 |\hat{Q}(\hat{Q}+\hbar\delta)|- \frac{1}{2}(\hat{K}\hat{K}^{\dagger}+
 \hat{K}^{\dagger}\hat{K})= \frac{1}{2}[\hat{K},\hat{K}^{\dagger}]\nonumber\\
&=&
 \left\{\begin{array}{cl}  \hbar\delta\hat{Q} & \mbox{if }|Q|\geq\hbar\delta\\
    \hat{Q}^2{\rm sgn}\hat{Q} & \mbox{if
    }|Q|<\hbar\delta \end{array}\right.\,.
\end{eqnarray}
Therefore, restricted to the subspace on which $[\hat{K},\hat{K}^{\dagger}]$
is linear in $\hat{Q}$, we have an ${\rm sl}(2,{\mathbb R})$-representation
with Casimir operator
\begin{equation} \label{CasimirK}
 \hat{Q}^2-({\rm Re}\hat{K})^2-({\rm Im}\hat{K})^2=0\,,
\end{equation}
which is inequivalent to the representation given by (\ref{CasimirJ}). This
representation, chosen implicitly in \cite{ACS}, is also a reducible
combination of discrete-series representations, given by $D_1^+\oplus D_1^-$.

If one uses a representation in which $\hat{Q}$ has discrete spectrum
$\hbar\delta{\mathbb Z}$, only a 1-dimensional subspace is subject to the
formally non-linear relation in (\ref{KK}). Since this subspace is annihilated
by $\hat{Q}$, there is in fact no non-linearity in such a
representation. However, the non-linearity exhibited in (\ref{KK}) may be
relevant for other representations, for instance if one follows \cite{Bohr}
and uses a non-separable Hilbert space on which $\hat{Q}$ has a discrete
spectrum containing all real numbers, including non-zero eigenvalues in the
range between $-\hbar\delta$ and $\hbar\delta$.  Even then, the deviation from
linear behavior is rather weak, and affects only wave functions that have a
small expectation value $|\langle\hat{Q}\rangle|<\hbar\delta$. For such
solutions, which are relevant if infrared renormalization is taken into
account, quantum back-reaction of fluctuations on expectation values is more
pronounced than in the complete ${\rm sl}(2,{\mathbb R})$-model
(\ref{Hdelta1}), and any identities between expectation values and moments
derived from properties of ${\rm sl}(2,{\mathbb R})$ may be violated. The
latter consequence may have implications for non-bouncing solutions, to which
we turn now.

\section{Solutions}

Both models, the original one using the Hamiltonian (\ref{Hdelta1}) in
\cite{BouncePert} and the later one using the Hamiltonian (\ref{Hdelta2}) in
\cite{ACS} have been called ``solvable.'' However, the degree of solvability
as demonstrated so far is quite different. Since (\ref{Hdelta1}) is part of a
completely linear system, there is no quantum back-reaction of fluctuations
and higher moments of a state on expectation values, leading to dynamical
coherent behavior. Using (\ref{Hdelta2}), it so happens that one can find
analytic solutions for evolving wave functions, but this is a formal kind of
solvability that does not need to imply any physical properties. In fact, as
shown by (\ref{KK}), this choice of factor orderings may lead to non-linear
brackets and therefore quantum back-reaction, depending on how $\hat{Q}$ is
represented. Dynamical states that get close to small volume in this model do
not fully maintain coherence. However, if one uses a representation on which
$\hat{Q}$ has discrete spectrum $\hbar\delta{\mathbb Z}$, our result
(\ref{KK}) demonstrates that the model of \cite{ACS} has a hidden form of
solvability equivalent to the explicit solvability of the model given in
\cite{BouncePert}. 

Even in the ordering (\ref{Hdelta1}) or in a restriction of (\ref{Hdelta2}) to
the subspace on which (\ref{KK}) is linear, the solvable nature of the
dynamics could be challenged by the absolute value in (\ref{Hdelta}), which is
not a linear function. However, as pointed out already in
\cite{BounceCohStates}, one can eliminate the absolute value if one works with
states that are supported only on either the positive or negative part of the
spectrum of ${\rm Im}\hat{J}$. For such states, the system with the ordering
(\ref{Hdelta1}) is fully linear. One may wonder whether restricting the
support of states could limit the size of quantum fluctuations of $Q$ or $P$,
which will play an important role in our subsequent classification of bouncing
and non-bouncing solutions. A separate paper will demonstrate that this is not
the case \cite{NonBouncingStates}.

\subsection{Fluctuations}

Given a linear algebra of basic operators $(\hat{Q},\hat{J})$ and the
Hamiltonian $\hat{H}_{\delta}$ in (\ref{Hdelta1}), Heisenberg's equations of
motion for operators or Ehrenfest's equations for expectation values are
linear:
\begin{eqnarray} \label{EOM}
 \frac{{\rm d}\langle\hat{Q}\rangle}{{\rm d}\lambda} &=& {\rm Re}
 \langle\hat{J}\rangle\\
 \frac{{\rm d}{\rm Re}\langle\hat{J}\rangle}{{\rm d}\lambda} &=&
 \langle\hat{Q}\rangle-\frac{1}{2}\hbar\delta\\
 \frac{{\rm d}{\rm Im}\langle\hat{J}\rangle}{{\rm d}\lambda} &=&0\,.
\end{eqnarray}
(The shift by $-\frac{1}{2}\hbar\delta$ is absent if (\ref{Hdelta2}) is
used as Hamiltonian.)  They can easily be solved by
\begin{equation} \label{QJ}
 \langle\hat{Q}\rangle(\lambda)= \frac{1}{2}\hbar\delta+
 \frac{1}{2}A\exp(\lambda)+\frac{1}{2}B\exp(-\lambda)\quad,\quad 
 {\rm Re}\langle\hat{J}\rangle(\lambda)=
 \frac{1}{2}A\exp(\lambda)-\frac{1}{2}B\exp(-\lambda) 
\end{equation}
with two constants $A$ and $B$, while ${\rm
  Im}\langle\hat{J}\rangle(\lambda)=\delta p_{\lambda}$ is constant.

The expectation values are subject to a quantum version of the classical
reality condition, which selects the Casimir of ${\rm sl}(2,{\mathbb
  R})$. Taking an expectation value of the identity (\ref{RealityJ}), we
obtain the condition
\begin{equation}
 \left(\langle\hat{Q}\rangle-\frac{1}{2}\hbar\delta\right)^2+(\Delta
 Q)^2-({\rm Re}\langle\hat{J}\rangle)^2- 
 (\Delta {\rm Re}J)^2 - ({\rm
   Im}\langle\hat{J}\rangle)^2- (\Delta{\rm
   Im}J)^2=-\frac{1}{4}\hbar^2\delta^2\,, 
\end{equation}
or
\begin{equation} \label{RealityExp}
  \left(\langle\hat{Q}\rangle-\frac{1}{2}\hbar\delta\right)^2-({\rm
    Re}\langle\hat{J}\rangle)^2= 
  \delta^2\left(p_{\lambda}^2-\hbar^2/4\right)+\delta^2(\Delta p_{\lambda})^2+
  (\Delta 
  {\rm Re}J)^2 - (\Delta Q)^2\,.
\end{equation}
(If (\ref{Hdelta2}) and (\ref{CasimirK}) are used, the latter condition reads
\begin{equation} \label{RealityExpK}
  \langle\hat{Q}\rangle^2-({\rm
    Re}\langle\hat{J}\rangle)^2= 
  \delta^2p_{\lambda}^2+\delta^2(\Delta p_{\lambda})^2+
  (\Delta 
  {\rm Re}J)^2 - (\Delta Q)^2\,.
\end{equation}
The shift of $\langle\hat{Q}\rangle$ by $-\frac{1}{2}\hbar\delta$ on the
left-hand side of (\ref{RealityExp}), as well as the negative term
$-\frac{1}{4}\hbar^2\delta^2$ on the right-hand side of this equation, are
subject to the main two quantization ambiguities, as we will see in
Sec.~\ref{s:AmbigRep}.)

For fluctuations $\Delta Q$ smaller than $\delta
\sqrt{p_{\lambda}-\hbar^2/4}$, the right-hand side of (\ref{RealityExp}) is
guaranteed to be positive. The resulting condition
$(\langle\hat{Q}\rangle-\frac{1}{2}\hbar\delta)^2-({\rm
  Re}\langle\hat{J}\rangle)^2=AB>0$ can then be fulfilled for the
$\lambda$-dependent solutions only if $AB>0$ in (\ref{QJ}). By adjusting the
zero value of $\lambda$, we can always assume that $B=|A|$ unless $B=0$ or
$A=0$. For $AB>0$, $B=A>0$, such that
\begin{equation} \label{Qcosh}
 \langle \hat{Q}\rangle(\lambda)= \frac{1}{2}\hbar\delta +A \cosh(\lambda)
\end{equation}
follows a bouncing trajectory. 

In order to show that a bounce happens generically in this model, one should
demonstrate that $Q$-fluctuations can never be so large that the right-hand
side of (\ref{RealityExp}) is no longer positive. Semiclassical situations
cannot lead to a non-bouncing scenario because semiclassical fluctuations
$\Delta Q$ cannot overcome the large $\delta p_{\lambda}$. However, as we
approach a BKL-type singularity, infrared renormalization implies that the
averaging volume $V_0$ gets smaller and smaller, without any non-zero lower
bound in the classical theory. Moreover, since the energy density
$p_{\lambda}^2/2V^2$ is independent of $V_0$, $p_{\lambda}$ decreases with
decreasing $V_0$. Since $p_{\lambda}$ decreases by a classical effect,
unrelated to $\hbar$, it is not inconceivable that
$\delta^2(p_{\lambda}^2-\hbar^2/4)$ could be smaller than
\begin{equation} \label{Delta}
 \Delta=(\Delta Q)^2-\delta^2(\Delta p_{\lambda})^2- (\Delta {\rm
  Re}J)^2\,.
\end{equation}
Note that products of fluctuations such as $\Delta Q\Delta P$ are bounded from
below by uncertainty relations independent of $V_0$. Therefore, they are not
suppressed as much by infrared renormalization and can remain significant even
as $Q$ gets smaller and smaller; see \cite{EFTLQC} for a detailed discussion.

If the fluctuations collected in $\Delta$ are so large that they cancel out
the term $\delta^2(p_{\lambda}^2-\hbar^2/4)$, the constants in our solutions
(\ref{QJ}) have to obey $AB=0$, such that $A=0$ or $B=0$, in which case we
have a non-bouncing
\begin{equation} \label{Qexp}
 \langle\hat{Q}\rangle(\lambda)=\frac{1}{2}\hbar\delta+
 \frac{1}{2}A\exp(\pm\lambda) 
\end{equation}
that resembles the singular classical solutions. If fluctuations are just
slightly larger, $AB<0$ implies that we can adjust the zero value of $\lambda$
such that $B=-A$, in which case
\begin{equation}\label{Qsinh}
 \langle\hat{Q}\rangle(\lambda)= \frac{1}{2}\hbar\delta \pm A\sinh(\lambda)
\end{equation}
follows a non-classical non-bouncing trajectory. 

The relationship between fluctuations and expectation values, or detailed
knowledge of the quantum state close to a singularity, is therefore required
to see whether a bounce happens generically. The non-trivial nature of this
behavior is shown by a result of \cite{FluctEn}: For a state Gaussian in $Q$,
$\Delta$ is always negative, even if the state is squeezed. Expectation values
of such a state always bounce, provided that
$p_{\lambda}>\frac{1}{2}\hbar$. For a state which is still Gaussian and
possibly squeezed, but in $\log |Q|$ rather than $Q$, $\Delta=0$ and only the
small (infrared-renormalized) $\delta^2 p_{\lambda}^2$ remains in addition to
$-\hbar^2/4$. This contribution may still be positive, so that these states
could bounce as well, but only if $p_{\lambda}$ is above a minimal value which
need not be respected by infrared renormalization. Since the model of
\cite{ACS} implies the condition (\ref{RealityExpK}) in which the term
$-\frac{1}{4}\hbar^2\delta^2$ is absent, any non-zero $p_{\lambda}$, however
small, would lead to bouncing solutions even if $\Delta=0$. In this sense,
\cite{ACS}, compared with \cite{BouncePert}, makes bouncing solutions more
likely. 

So far the possibility has not been ruled out that a non-Gaussian state might
lead to $\Delta>0$, such that even the positive $\delta^2 p_{\lambda}^2$ could
be overcome. No such state has been found yet, but not much of the state space
has been explored beyond Gaussian ones, and going beyond Gaussian wave
functions in a systematic way requires a tedious analysis. At present it is
therefore impossible to conclude, based on such methods, that a bounce happens
generically. However, we are now able to demonstrate the genericness of
bouncing solutions within the model of \cite{ACS} (but not within loop quantum
cosmology in general) using our identification of this model with the
reducible representation of ${\rm sl}(2,{\mathbb R})$ derived in
Sec.~\ref{s:Rep}.

\subsection{Representation theory}
\label{s:BounceRep}

There is an interesting relationship between the possibility of non-bouncing
solutions and representation theory of ${\rm sl}(2,{\mathbb R})$. In the two
examples with Casimirs (\ref{CasimirJ}) and (\ref{CasimirK}), respectively, it
is possible to use a reducible representation which is a direct sum of two
irreducible ones, one such that $Q>0$ and one such that $Q<0$ in terms of
$\hat{Q}$-eigenvalues. Therefore, an evolving state supported on $Q>0$
initially will always be supported on $Q>0$ if evolution is generated by an
${\rm sl}(2,{\mathbb R})$-element in the same representation. Time-dependent
expectation values such as (\ref{Qsinh}) are then impossible, and $Q=0$ will
never be crossed. The non-bouncing possibility (\ref{Qexp}) with $Q=0$
approached asymptotically is not ruled out by this statement. However, such a
solution, with $AB=0$ in (\ref{QJ}), requires a specific value of the
fluctuation parameter $\Delta$ in (\ref{Delta}) for fixed $p_{\lambda}$, and
would therefore not be considered generic. Within a given model of this
form, bouncing solutions are therefore generic, but we should examine the
possibility of other ${\rm sl}(2,{\mathbb R})$-representations, in addition to
what has serendipitously been selected in \cite{BouncePert} or \cite{ACS},
before we can tell whether bouncing solutions are generic within loop quantum
cosmology, understood as any quantum model with holonomy-modified dynamics.

The reducible representations implicitly used in \cite{BouncePert} and
\cite{ACS} make use of the discrete series, on which the Casimir $R$ has to
respect the inequality $R\geq -\frac{1}{4}\hbar^2\delta^2$
\cite{Bargmann}. The limiting value is realized in (\ref{CasimirJ}), while
(\ref{CasimirK}) makes a more advantageous choice of the Casimir, increasing
the likelihood of bouncing solutions. For general $R$,
$\delta^2(p_{\lambda}^2-\hbar^2/4)$ in (\ref{RealityExp}) is replaced by
$\delta^2p_{\lambda}^2+R$. (Moreover, as shown in (\ref{RealityExpK}), the
shift of $\langle\hat{Q}\rangle$ by $\frac{1}{2}\hbar\delta$ in
(\ref{RealityExp}) is absent if the derivation is repeated in the model of
\cite{ACS} where according to (\ref{KK}) an unshifted $\hat{Q}$ is one of the
algebra generators.) Compared with the limiting value of (\ref{CasimirJ}),
larger fluctuations are therefore necessary for $AB=0$ in (\ref{QJ}) to result
from imposing the reality condition, and non-bouncing solutions are less
likely. Choosing $R=0$, as implicitly done in \cite{ACS}, therefore makes
bounces more likely than choosing $R=-\frac{1}{4}\hbar^2\delta^2$ as in
\cite{BouncePert}, and representations with $R>0$ that have not been studied
yet would further enhance this likelihood.

Representations in the continuous series, which so far have not been studied
in this context either, respect the inequality $R< 0$, all of which decrease
the likelihood of a bounce. Moreover, in this case a representation containing
both positive and negative eigenvalues of $Q$ is irreducible, such that a
Hamiltonian constructed in such a representation is able to map a state
supported on $Q>0$ to a state with some support on $Q<0$. The non-bouncing
possibility of (\ref{Qsinh}) is therefore not ruled out, and bouncing
solutions do not appear to be generic. Our argument using fluctuations
$\Delta$ in (\ref{RealityExp}) shows that non-bouncing solutions should indeed
be more likely precisely for a negative range of $R$, in which case we have
irreducible representations in which $Q>0$ can be mapped to $Q<0$. From a
fundamental perspective, see for instance \cite{Isham}, one may prefer an
irreducible representation of the dynamical algebra, which supports the
possibility of non-bouncing solutions.

\section{Analog harmonic oscillators}
\label{s:Analog}

The role of fluctuations can nicely be illustrated by analog models using
upside-down harmonic oscillators. The classical Hamiltonian $QP$ is equivalent
to such an oscillator based on the canonical transformation $\sqrt{2}\;q=Q-P$,
$\sqrt{2}\;p=Q+P$ such that $QP=\frac{1}{2}(p^2-q^2)$.

However, for a generalization to the holonomy-modified model, it is more
convenient to proceed without applying a canonical transformation. In the
classical case, we can derive an analog model of the Hamiltonian system
generated by $H=QP$ by rewriting the first-order equations it generates,
$\dot{Q}=Q$ and $\dot{P}=-P$, in terms of a second-order equation for $Q$. If
$\dot{Q}=Q$, we have $\ddot{Q}=\dot{Q}=Q$, which is the second-order equation
of motion generated by the upside-down harmonic Hamiltonian
\begin{equation} \label{Hanalog}
 H_{\rm analog}=\frac{1}{2}(\pi^2-Q^2)
\end{equation}
with momentum $\pi=\dot{Q}$. Again using $\dot{Q}=Q$, we obtain $\pi=Q$ or
$E=H_{\rm analog}=0$ on solutions we are interested in. The first-order
equations generated by $QP$ are therefore equivalent to equations of motion
generated by $H_{\rm analog}$ together with the condition that the energy be
zero. Standard knowledge about the upside-down oscillator then immediately
shows that we have solutions of the form (\ref{Qexp}) in which $Q=0$ is
approached asymptotically.

For the loop model, the first-order equations generated by (\ref{Hdelta1}) are
given in (\ref{EOM}). Ignoring the constant shift of $Q$ by
$-\frac{1}{2}\hbar\delta$, which is irrelevant here, we still have the
second-order equation $\ddot{Q}={\rm d}{\rm Re}J/{\rm d}\lambda=Q$,
corresponding to the same analog Hamiltonian $H_{\rm analog}$ used for the
unmodified dynamics in (\ref{Hanalog}). However, $\pi=\dot{Q}={\rm Re}J\not=Q$
in general, such that we are looking for solutions with non-zero
energy. Again, standard knowledge about the upside-down oscillator shows that
negative energies give rise to bouncing solutions (\ref{Qcosh}), while
positive energies imply non-bouncing solutions (\ref{Qsinh}).

In the loop model, the energy of analog solutions is determined by quantum
fluctuations. Since $\pi={\rm Re}J$, we can derive its relationship with $Q$
using the reality condition 
\begin{equation}
 -2E=Q^2-({\rm Re}J)^2=\delta^2(p_{\lambda}^2-\hbar^2/4)-\Delta\,.
\end{equation}
(See (\ref{RealityExp}) and (\ref{Delta}).)
Therefore, we have $E<0$ and bouncing solutions if $\Delta$ is sufficiently
small or negative. However, we may have non-bouncing solutions of the form
(\ref{Qexp}) if $E=0$, or of the form (\ref{Qsinh}) if $E>0$. In the latter two
cases, $p_{\lambda}$ must be such that $\delta^2(p_{\lambda}^2-\hbar^2/4)<\Delta$.

Factor-ordering choices affect the potential of an analog model. For instance,
the non-linear relation (\ref{KK}) for $|Q|<\hbar\delta$ implies that
$\ddot{Q}$ is proportional to $Q^2{\rm sgn}Q$ instead of $Q$, which requires
an anharmonic potential proportional to $-|Q|^3$. This potential still
vanishes at $Q=0$ where it has a local maximum. Qualitatively, the behavior of
solutions with different energies is therefore similar to the harmonic analog
system, but the anharmonic potential likely leads to more significant changes
of fluctuations, or $\Delta$, during evolution.

In these analog models, the question of whether there is a bounce is a matter
of initial values rather than the dynamics. Initial values, in turn, are
determined by the quantum state relevant in the small-volume regime.

\section{Bounded densities}
\label{s:Bounds}

Different versions of upper bounds on the energy density have been derived in
Hilbert-space representations of models of loop quantum cosmology
\cite{ACS,DensityOp}. At first sight, the possibility of non-bouncing
solutions for $Q$ approaching or crossing zero seems to be inconsistent with
such bounds, a conclusion which is often suggested in the literature. However,
we can explicitly demonstrate that there is no inconsistency. After all, the
density $\widehat{p_{\phi}^2/2V^2}$, or as a simpler substitute the {\em
  positive} expressions $\hat{Q}^2$ or $\hat{Q}^{-2}$, may have bounded
expectation values even if $\langle\hat{Q}\rangle$ is zero, simply because
$\langle\hat{Q}^2\rangle=\langle\hat{Q}\rangle^2+(\Delta Q)^2$ contains a
contribution from fluctuations.

A more refined argument that suggests a strict relationship between density
bounds and bouncing solutions refers to the Planckian value of the upper bound
on energy densities derived in \cite{ACS,DensityOp}, which agrees with the
Planckian density usually obtained at the bounce point of bouncing solutions
in models of loop quantum cosmology. However, even such a quantitative
relationship does not imply that general statements about energy bounds imply
bouncing solutions. In order to demonstrate this perhaps subtle statement, we
use the algebraic model to derive a bound on $\langle\hat{Q}^2\rangle$, which
we can then analyze in the bouncing case $AB>0$ in (\ref{QJ}) and in the
non-bouncing one, $AB\leq 0$. To be specific, we first assume ${\rm
  sl}(2,{\mathbb R})$-relations such that there is no shift of $Q$ and a zero
Casimir, corresponding to the model of \cite{ACS}.

Ehrenfest's equations of motion can be derived in the linear model not only
for expectation values but also for fluctuations. Unlike expectation values,
fluctuations are subject to uncertainty relations, which bound possible
initial values for their equations of motion.  Fluctuations and expectation
values therefore have different positivity properties, which is relevant for
the existence of local minima. 

Volume fluctuations in the linear model always obey the relation
\cite{BounceCohStates}
\begin{equation} \label{DeltaQ}
  (\Delta Q)^2(\lambda)=
  \frac{1}{2}\left(c_3\exp(-2\lambda)+c_4\exp(2\lambda)\right)- 
  \frac{1}{4}(c_1+c_2)
\end{equation}
with constants $c_i$, where
\begin{equation} \label{c1}
 c_1=-\Delta= AB-\delta^2p_{\lambda}^2
\end{equation}
and
\begin{equation}\label{c1c2}
 c_1-c_2= 2\delta^2 (\Delta p_{\lambda})^2\,.
\end{equation}
Since $(\Delta Q)^2(\lambda)\geq 0$ for all $\lambda$, $c_3$ and $c_4$ cannot
be negative. Moreover, as shown in \cite{Harmonic}, uncertainty relations
imply that
\begin{equation}
 c_3c_4\geq \hbar^2\delta^2p_{\lambda}^2+ \frac{1}{4}(c_1+c_2)^2> 0\,.
\end{equation}
Therefore, $c_3$ and $c_4$ are strictly positive for all states, such that
$(\Delta Q)^2(\lambda)$ always has a local minimum. It is located at
$\lambda=\frac{1}{4}\log (c_3/c_4)$, at which time we have the minimal
fluctuations
\begin{equation}
 (\Delta Q)^2_{\rm min}= \sqrt{c_3c_4}-\frac{1}{4}(c_1+c_2)\,.
\end{equation}
For a dynamical coherent state \cite{BounceCohStates}, for instance,
\begin{equation}
(\Delta Q)^2_{\rm min}=
 \frac{1}{2}\sqrt{\hbar^2\delta^2p_{\lambda}^2+ \delta^4 (\Delta
   p_{\lambda})^4+ \Delta^2+2\Delta\delta^2 (\Delta p_{\lambda})^2}+
 \frac{1}{2}\delta^2(\Delta p_{\lambda})^2+ \frac{1}{2}\Delta
\end{equation}
in which $\Delta$ is the only parameter characterizing the state that does not
refer to the matter ingredients. It is easy to see that $(\Delta Q)^2_{\rm
  min}$, as a function of $\Delta$, is monotonically increasing, with
$\lim_{\Delta\to\infty} (\Delta Q)^2_{\rm min}=\infty$ and
$\lim_{\Delta\to-\infty} (\Delta Q)^2_{\rm min}=0$. For large $\Delta>0$, we
have the set of states that may not bounce because $\Delta$ can overcome
$\delta^2p_{\lambda}^2$. However, in this range of $\Delta$, $(\Delta
Q)^2_{\rm min}$ is large, which explains why
$\langle\hat{Q}^2\rangle=\langle\hat{Q}\rangle^2+(\Delta Q)^2$ can be bounded
even if $\langle\hat{Q}\rangle$ approaches or crosses zero.

In order to demonstrate a strict bound for $\langle\hat{Q}^2\rangle$ as well
as its Planckian nature, we combine $\langle\hat{Q}\rangle(\lambda)$ from
(\ref{QJ}) with $(\Delta Q)^2(\lambda)$ from (\ref{DeltaQ}):
\begin{eqnarray}
 \langle\hat{Q}^2\rangle(\lambda) &=& \langle\hat{Q}\rangle(\lambda)^2+ (\Delta
 Q)^2(\lambda)\nonumber\\
 &=& \frac{1}{4} \left((2c_3+A^2)\exp(-2\lambda)+
   (2c_4+B^2)\exp(2\lambda)\right)+ \frac{1}{2}AB- \frac{1}{4}(c_1+c_2)\,.
\end{eqnarray}
Because both $c_3$ and $c_4$ are positive, we have $(2c_3+A^2)(2c_4+B^2)>0$
and $\langle\hat{Q}^2\rangle(\lambda)$, unlike
$\langle\hat{Q}\rangle(\lambda)$, always has a local minimum, even if $AB\leq
0$:
\begin{equation}
 \langle\hat{Q}^2\rangle_{\rm min}= \frac{1}{2}\sqrt{(2c_3+A^2)(2c_4+B^2)}+
 \frac{1}{2}AB- \frac{1}{4}(c_1+c_2)\,.
\end{equation}
Since the square root is always positive, this minimum is bounded from below
by
\begin{equation} \label{Q2min}
 \langle\hat{Q}^2\rangle_{\rm min} >
 \frac{1}{2}AB- \frac{1}{4}(c_1+c_2) = \frac{1}{2}\delta^2p_{\lambda}^2+
 \frac{1}{4}(c_1-c_2)= \frac{1}{2}\delta^2\left(p_{\lambda}^2+(\Delta
 p_{\lambda})^2\right)=\frac{1}{2}\delta^2\langle\hat{p}_{\lambda}^2\rangle\,,
\end{equation}
using (\ref{c1}) and (\ref{c1c2}).

We are now ready to obtain an upper bound for the energy density
$\rho=p_{\phi}^2/2V^2$, which we define as
$\langle\hat{p}_{\phi}^2\rangle^2/2\langle\hat{V}^2\rangle$ on quantum states.
(There is no unique operator for the energy density. Other definitions are
related to our choice through additional fluctuation terms, which may change
upper bounds.) If $\langle\hat{Q}^2\rangle$ is bounded from below, $\rho$ is
bounded from above for fixed $p_{\phi}$. Using the relationships $|Q|=4\pi GV$
and $p_{\phi}=\sqrt{12\pi G}p_{\lambda}$ for $x=-1/2$, we obtain
\begin{equation}
 \rho=\frac{3}{8\pi G}\frac{p_{\lambda}^2}{Q^2}\leq \frac{3}{4\pi G\delta^2}=2
 \rho_{\rm QG}
\end{equation}
where $\rho_{\rm QG}$ is the energy density at which large-volume solutions
(\ref{Qcosh}) bounce, such that $\rho_{\rm QG}\approx \rho_{\rm P}$ if
$\delta\approx \ell_{\rm P}$ for $x=-1/2$. Therefore, even if we do not have a
bouncing solution, that is if $A$ and $B$ are such that the volume is of the
form (\ref{Qsinh}) or (\ref{Qexp}), the energy density is bounded from above
by a fixed multiple of the Planck density. Similarly, for general $x$ we have
the $Q$-dependent density bound
\begin{eqnarray}
 \rho&=&\frac{3}{8\pi G}\frac{p_{\lambda}^2}{Q^2 ((8\pi G/3)
   (1-x)|Q|)^{(1+2x)/(1-x)}}\nonumber\\
& \leq& \frac{3}{4\pi G\delta^2((8\pi G/3)
   (1-x)|Q|)^{(1+2x)/(1-x)}}=\frac{2 \rho_{\rm QG}}{((8\pi G/3)
   (1-x)|Q|)^{(1+2x)/(1-x)}}\,.
\end{eqnarray}

It is interesting to contrast the limiting case $AB=0$ with the
Wheeler--DeWitt model, that is, the quantized model without holonomy
modifications. For $\hat{H}=\pm\frac{1}{2}(\hat{Q}\hat{P}+\hat{P}\hat{Q})$,
Ehrenfest's equation
\begin{equation}
 \frac{{\rm d}\langle\hat{Q}\rangle}{{\rm
     d}\lambda}=\pm\langle\hat{Q}\rangle
\end{equation}
is solved by $\langle\hat{Q}\rangle(\lambda)=\frac{1}{2}A\exp(\pm\lambda)$,
which is of the form (\ref{Qexp}). However, fluctuations in this model are
crucially different from (\ref{DeltaQ}): For fluctuations, Ehrenfest's
equations imply
\begin{equation}
 \frac{{\rm d}(\Delta Q)^2}{{\rm d}\lambda}= \pm2(\Delta Q)^2\,,
\end{equation}
such that $\Delta Q\propto \exp(\pm\lambda)$. Although this solution (squared)
is formally of the form (\ref{DeltaQ}), it would require $c_3c_4=0$ which in
the holonomy-modified model is ruled out by uncertainty relations. For this
reason, the Wheeler--DeWitt model does not obey density bounds: Our derivation
cannot be applied to this model because $(\Delta Q)^2(\lambda)$ then does not
have a local minimum. The solution (\ref{Qexp}), obtained with $AB=0$ in the
loop model, therefore describes a different state compared with the
Wheeler--DeWitt solution, even though it has an identical behavior of
$\langle\hat{Q}\rangle(\lambda)$.

\section{Quantization ambiguities}

We have already seen one example of a factor ordering choice, using $\hat{K}$
in (\ref{KK}) instead of $\hat{J}$ as implicitly done in \cite{ACS}, that
affects the dynamics of states, in particular at small volume. Such
ambiguities are therefore relevant for the question of whether there may be a
generic bounce. Other ordering ambiguity can be formulated within the
completely linear model.

\subsection{Representations}
\label{s:AmbigRep}

Quantizing the solvable model amounts to the choice of an irreducible
representation of ${\rm sl}(2,{\mathbb R})$. Inequivalent representations are
classified by the value $R$ of the Casimir operator $\hat{Q}^2-({\rm
  Re}\hat{J})^2- ({\rm Im}\hat{J})^2=R$. As we have seen, the interpretation
$|{\rm Im}J|=\delta p_{\lambda}$ in a cosmological model means that
expectation values of the volume variable $Q$ and its time derivative related
to ${\rm Re}J$ obey the relation
\begin{equation}
 \langle\hat{Q}\rangle^2-\langle{\rm Re}\hat{J}\rangle^2=
 \delta^2p_{\lambda}^2+R-\Delta
\end{equation}
with a fluctuation term $\Delta$; see (\ref{RealityExp}). If the right-hand
side is positive, $\langle\hat{Q}\rangle(\lambda)$ is cosh-like and bounces,
while it is exponential or sinh-like, and therefore non-bouncing, if the
right-hand side is zero or positive.

For unitary irreducible representations of ${\rm sl}(2,{\mathbb R})$ on which
$\hat{Q}$ has positive and negative eigenvalues, we must use the continuous
series, on which the Casimir is restricted by the inequality $R< 0$. Bouncing
solutions are most likely for small $|R|$, close to the discrete-series
value implicitly chosen in the quantization used in \cite{ACS}. But any
negative $R$ helps to overcome the positive term $\delta^2p_{\lambda}^2$,
which is small thanks to infrared renormalization. A quantization using $R$
close to zero, as in \cite{ACS}, is therefore highly non-generic, and it
cannot be used to determine robust features of possible quantum models.

In terms of reducible representations with both positive and negative
eigenvalues of $\hat{Q}$, there are infinitely many choices in addition to the
two explicit examples (\ref{CasimirJ}) and (\ref{CasimirK}) shown so far. For
any $k>0$, we may use the representation $D_k^+\oplus D_k^-$ such that
$R=k(k-1)$. (Recall that $k=1/2$ in (\ref{CasimirJ}) and $k=1$ in
(\ref{CasimirK}).) On these representations, the eigenvalues of $Q$ are given
by $|Q|\in k+{\mathbb N}_0$ and therefore obey the inequality $|Q|\geq k$.
All these representations imply Hamiltonians that preserve ${\rm sgn}Q$ and
therefore rule out the non-bouncing behavior (\ref{Qsinh}), but not
(\ref{Qexp}).  Any choice with $k\geq 1$ makes bouncing solutions more likely
because the fluctuation term $\Delta$ would have to overcome not only
$\delta^2p_{\lambda}^2$ but $\delta^2p_{\lambda}^2+k(k-1)$ for solutions with
$AB=0$ in (\ref{QJ}). However, choosing these representations, which are
reducible and such that a range of $Q$-eigenvalues around zero is excluded,
would constitute an ad-hoc way of increasing the likelihood of bouncing
solutions.

The value of $R$ also has an effect on density bounds. If $R\not=0$,
$\delta^2p_{\lambda}^2$ in (\ref{c1}), and therefore
$\delta^2\langle\hat{p}_{\lambda}^2\rangle$ in (\ref{Q2min}), is replaced by
$\delta^2\langle\hat{p}_{\lambda}^2\rangle+R$. For negative $R$,
$\langle\hat{Q}^2\rangle_{\rm min}$ may be zero for some states, such that no
upper bound on the expected density is then obtained even if $p_{\lambda}$ is
fixed. Also regarding the question of bounded densities, therefore, \cite{ACS}
inadvertently made a beneficial but highly non-generic choice of a
quantization.

In order to see another ambiguity, let us define the ${\rm sl}(2,{\mathbb R})$
ladder operator $\hat{J}$ as
\begin{equation}
 \hat{J}=\hat{h}_{\delta_1}\hat{Q}\hat{h}_{\delta_2}
\end{equation}
with $\delta_1+\delta_2=\delta$, such that our previous $\hat{J}$ is obtained
for $\delta_1=0$. Using basic commutator relationships as before, we derive
\begin{equation}
 \hat{J}= (\hat{Q}+\hbar\delta_1)\hat{h}_{\delta}\,.
\end{equation}
The brackets $[\hat{Q},\hat{J}]$ and
$[\hat{Q},\hat{J}^{\dagger}]$ remain unchanged, but we now have
\begin{equation}
 [\hat{J},\hat{J}^{\dagger}]= 2\hbar\delta
 \left(\hat{Q}+\hbar(\delta_1-\delta/2)\right)\,.
\end{equation}
The Casimir $R=-\frac{1}{4}\hbar^2\delta^2$ in this class of quantizations is
independent of $\delta_1$.

The linear nature of the model is preserved, but in our previous relations
$\hat{Q}$ is shifted by a constant to $\hat{Q}+\hbar\delta_1$. In particular,
an arbitrary constant $\hbar\delta_1$ can be subtracted from a bouncing
solution (\ref{Qcosh}), possibly pushing the local minimum to negative values,
accompanied by two zeros of $\langle\hat{Q}\rangle(\lambda)$. For small-volume
solutions (after infrared renormalization), even a small $\hbar\delta_1$ can
lead to this singular behavior. Large fluctuations are not required in this
case. With a positive $\delta_1$, this type of quantization ambiguity can lead
to singular solutions even in quantizations based on reducible representations
using the discrete series: Even though the expectation value of the
corresponding ${\rm sl}(2,{\mathbb R})$-generator
$\hat{L}_0=\hat{Q}+\hbar\delta_1$ would never cross zero based on our previous
arguments, the expectation value of the triad operator
$\hat{Q}=\hat{L}_0-\hbar\delta_1$ can cross zero if $\delta p_{\lambda}$ is
sufficiently small.

\subsection{Inverse-volume corrections}
\label{s:Inv}

Another source of quantization ambiguites is given by inverse-volume
corrections \cite{InvScale}, which we have not considered in detail so
far. These corrections do not preserve the linearity of the model and can
therefore lead to further differences compared with a simple bouncing solution
such as (\ref{Qcosh}). An obvious place for inverse-volume corrections to
appear is the matter Hamiltonian, $\frac{1}{2} p_{\phi}^2/V$ for a free
massless scalar. If this Hamiltonian is multiplied with a correction function
that approaches zero at small $V$, the small-volume behavior is clearly
modified. The general small-volume behavior of inverse-volume corrections is
of the power-law form $f(Q)\sim f_0 Q^n$ with a positive integer $n>0$, where
both $n$ and $f_0$ depend on quantization ambiguities \cite{Ambig,ICGC}.

A second, less obvious place for inverse-volume corrections is on the
gravitational side of the Hamiltonian constraint or the Friedmann equation
(\ref{Friedmann}). Even though the classical contribution $VP^2$ does not
contain an inverse $V$, any embedding of this isotropic term in an anisotropic
model requires inverse triad components \cite{HomCosmo,Spin}. Therefore, both
sides of the Friedmann equation can receive independent correction functions
of the form $f(Q)\sim f_0 Q^n$, with different $f_0$ and $n$ in each case. We
can then combine these two functions by bringing both of them to one side, say
the gravitational one. The ratio of these two functions is again of the form
$f(Q)\sim f_0 Q^n$, but the integer $n$ is no longer restricted to be
positive.

Inserting such a function in the full Hamiltonian constraint and solving for
$p_{\phi}$ implies that the previous Hamiltonian $QP$ is replaced by the
non-quadratic $|Q|^{1-n/2}P$, or the non-linear $|Q|^{-n/2} {\rm Im}J/\delta$. In
terms of analog models, $\ddot{Q}=(1-n/2)|Q|^{-n/2}\dot{Q}{\rm sgn}Q=
(1-n/2)|Q|^{1-n} {\rm sgn}Q$
requires a potential
\begin{equation}
 W(Q)= -\frac{1}{2} |Q|^{2-n}\,.
\end{equation}
For $n\leq 2$, the qualitative behavior of solutions for given energy values
is the same as in the harmonic case. If $n>0$, $Q=0$ may be reached even for
negative-energy solutions. Inverse-volume corrections therefore make it more
likely that small-volume solutions do not bounce.

\section{Implications for signature change}
\label{s:Sig}

The possibility of various bouncing or non-bouncing solutions (\ref{Qcosh}),
(\ref{Qexp}), or (\ref{Qsinh}) implies an interesting behavior regarding
signature change. This phenomenon has so far been considered only for bouncing
solutions (\ref{Qcosh}). But modified space-time structures that could give
rise to signature change are most likely realized at small volume where
non-bouncing solutions are possible as well.

Space-time structure cannot be derived within a homogeneous model but rather
requires an embedding in perturbative or some other form of inhomogeneity. The
perturbative \cite{ConstraintAlgebra,ScalarHolInv} and midisuperspace case
\cite{JR,HigherSpatial,GowdyCov} have been studied in some detail, indicating
a generic modification of the space-time structure as a consequence of
holonomy corrections \cite{DeformedCosmo}: Whenever an inhomogeneous
Hamiltonian constraint $H[N]$ is holonomy-modified by extending the
replacement of $P$ with $\sin(\delta P)/\delta$ to inhomogeneity, its Poisson
bracket
\begin{equation}
 \{H[N_1],H[N_2]\} = D[\beta(P) q^{ab}(N_1\partial_bN_2-N_2\partial_bN_1)]
\end{equation}
differs from the classical bracket ($\beta=1$) by a function
\begin{equation} \label{beta}
 \beta(P)=\cos(2\delta P)\,.
\end{equation}
Via Dirac's hypersurface deformations \cite{DiracHamGR}, any $\beta\not=1$
demonstrates a modified space-time structure. In particular, if $\beta$ can be
negative for certain $P$, such as around a local maximum of $\sin(\delta P)$
in the case of (\ref{beta}), space-time has Euclidean signature
\cite{Action,SigChange,SigImpl,Normal,EffLine}.

For bouncing solutions (\ref{Qcosh}), signature change happens around the
local minimum of $\langle\hat{Q}\rangle(\lambda)$, implying that the bounce,
even if it occurs, is not deterministic. (The initial-value problem is not
well-posed in Euclidean signature.) This result can be rederived for our
solutions, where, as a new feature, we take into account the quantization
ambiguity $\delta_1$. In order to express $\beta$ in terms of our solutions
(\ref{QJ}), we use ${\rm Re}J=Q\cos(\delta P)$ and write
\begin{equation}
 \beta=\cos(2\delta P)= 2\cos^2(\delta P)-1= 2\left(\frac{{\rm
       Re}J}{Q}\right)^2-1\,.
\end{equation}
The bouncing solution (\ref{Qcosh}) is obtained for $AB>0$, such that we can
assume $A=B>0$ by choosing a suitable zero value of $\lambda$. Therefore,
\begin{equation}
 {\rm Re}J(\lambda)=A\sinh(\lambda)\quad,\quad Q(\lambda)=
 A\cosh(\lambda)-\hbar\delta_1
\end{equation}
implies a non-constant
\begin{equation}
 \beta(\lambda)=
 1-2\frac{1-2\hbar\delta_1A^{-1}
   \cosh(\lambda)+\hbar^2\delta_1^2/A^2}{(\cosh(\lambda)-\hbar\delta_1/A)^2}
 \,. 
\end{equation}
While $\beta\to 1$ for large $\lambda$, $\beta(0)=-1$ at the local minimum of
$Q(\lambda)$. Around the bounce, space-time is therefore Euclidean.

The situation is rather different for our new, non-bouncing solutions. The
case of $AB=0$, or (\ref{Qexp}), is interesting because it implies that ${\rm
  Re}J=Q+\hbar\delta$. In this case, 
\begin{equation}
 \beta=2\left(1+\frac{\hbar\delta_1}{Q}\right)^2-1=
 1+4\frac{\hbar\delta_1}{Q}+ 2\frac{\hbar^2\delta_1^2}{Q^2}\,.
\end{equation}
Therefore, $\beta=1$ of $\delta_1=0$, and
the classical space-time structure is realized even at small volume. For
$AB<0$, we have
\begin{equation}
 {\rm Re}J(\lambda)=A\cosh(\lambda)\quad,\quad Q(\lambda)=
 A\sinh(\lambda)-\hbar\delta_1
\end{equation}
and
\begin{equation}
 \beta(\lambda)=
 1+2\frac{1+2\hbar\delta_1A^{-1}
   \sinh(\lambda)-\hbar^2\delta_1^2/A^2}{(\sinh(\lambda)-\hbar\delta_1/A)^2}=
 -1+2\frac{1+\sinh^2(\lambda)}{(\sinh(\lambda)-\hbar\delta_1/A)^2}   \,.
\end{equation}
This function is negative for $\sinh(\lambda)$ between
\begin{equation}
 s_{\pm}=-\frac{\hbar\delta_1}{A}\pm \sqrt{2}\sqrt{\hbar^2\delta_1^2/A^2-1}\,.
\end{equation}
For $\delta_1=0$, $\beta$ is always positive, such that there is no signature
change even though the space-time structure is non-classical ($\beta>1$).

\section{Conclusions}

We have derived several new results related to algebraic properties of
solvable models of loop quantum cosmology, relevant for the question of
whether bouncing solutions are generic. A copious amount of quantization
ambiguities has been illustrated by an explicit relationship between the
algebraic and Hilbert-space treatments of such models in Sec.~\ref{s:Rep}.

There are two main independent types of ambiguities, related to choices of
factor orderings and inequivalent representations through the value of a
Casimir variable. In Sec.~\ref{s:AmbigRep}, we have parameterized their
outcomes in solutions of dynamical equations by the shift $\delta_1$ of the
volume expectation value and the Casimir $R$ which appears in the reality
condition and determines implications of quantum fluctuations.  This result
has revealed that the Hilbert-space treatment given in \cite{ACS} is far from
being unique. Although the final representation (on a physical Hilbert space)
used in this context has been derived from a representation on a kinematical
Hilbert space, the latter is subject to choices and assumptions, for instance
regarding inner products, which are difficult to classify. The algebraic
treatment, by contrast, can build on the representation theory of ${\rm
  sl}(2,{\mathbb R})$ in order to determine possible quantization choices, and
to relate them to physical outcomes.  As a consequence, it can be seen that
\cite{ACS} implicitly made several specific choices for ambiguous objects that
increase the likelihood of bouncing solutions, corresponding to the values
$R=0$ and $\delta_1=0$ in our classification of Sec.~\ref{s:AmbigRep}.

Our analysis in Sec.~\ref{s:Rep} led us to an identification of the
quantization given in \cite{ACS} with a specific representation of ${\rm
  sl}(2,{\mathbb R})$. This representation is reducible, given by the direct
sum of two irreducible representations in the discrete series. Each of these
two irreducible representations is such that only eigenstates of the triad
operator $\hat{Q}$ of a specific sign are included. The representation used in
\cite{ACS} therefore does not include an operator that would map a state
supported on $Q>0$ to a state with some support on $Q<0$. This observation,
based on representation theory, allowed us to prove, for the first time, that
bouncing solutions are generic in the model constructed in \cite{ACS}; see
Sec.~\ref{s:BounceRep}. However, this conclusion has a sobering side too:
Within the set of all possible representations of ${\rm sl}(2,{\mathbb R})$,
which all amount to quantizations of holonomy-modified cosmological dynamics
and therefore constitute loop quantum cosmology, choosing a specific
representation, as implicitly done in the construction of \cite{ACS}, appears
to be rather ad-hoc.  Moreover, within the chosen representation, \cite{ACS}
also implicitly selected a factor ordering such that the triad operator
$\hat{Q}$ is a generator of ${\rm sl}(2,{\mathbb R})$. More generally, there
can be a constant shift determined by a quantization ambiguity $\delta_1$ in
Sec.~\ref{s:AmbigRep}, such that the generator $\hat{L}_0$ is given by
$\hat{Q}$ shifted by a constant. Each of the two irreducible representations
then has a fixed sign of $\hat{L}_0$-eigenvalues, and for a suitable sign of
the shift constant, $\langle\hat{Q}\rangle$ can cross zero and change sign even
if $\langle\hat{L}_0\rangle$ never does so in the given representation.

From this perspective, bouncing solutions are generic in the model of
\cite{ACS} only because the dynamics is formulated such that ${\rm sgn}Q$ is
fixed on an irreducible subrepresentation, which in combination with the
discrete spectrum of $\hat{Q}$ implies that $Q=0$ cannot be reached.  There
are irreducible representations of ${\rm sl}(2,{\mathbb R})$, each of which
includes states with positive as well as negative eigenvalues of
$\hat{Q}$. All representations in the continuous series are of this form, and
they may be preferred on fundamental grounds because they provide irreducible
representations of the dynamical algebra.  While \cite{ACS} made a
serendipitous choice of a representation that leads to generic bouncing
solutions, bounces are not generic within loop quantum cosmology in general.

While choices can be made, explicitly or implicitly, that increase the
likelihood of bouncing solutions, they still leave room for non-bouncing
solutions and therefore do not suffice to prove that bounces are generic in
loop quantum cosmology, in particular in the case of small-volume relevant
near a BKL-type singularity. Within the set of representations of ${\rm
  sl}(2,{\mathbb R})$, there is a large number of possibilities for
non-bouncing solutions which asymptotically approach or cross zero triad
expectation values. While evolution then remains meaningful in isotropic
models, the formulation of inhomogeneous cosmological modes on such degenerate
background geometries would be problematic. Throughout this discussion, we
have seen several detailed relationships between the representation theory of
${\rm sl}(2,{\mathbb R})$ and moment equations derived using canonical
effective theory.

In Sec.~\ref{s:Bounds}, we have shown that, perhaps surprisingly, non-bouncing
solutions may be consistent with Planckian upper bounds on energy densities,
but again the specific outcome depends on quantization ambiguities. Even if
the triad expectation value crosses zero, quantum fluctuations prevent the
volume expectation value from being zero. As a consequence, such upper bounds
strongly depend on the form of relevant quantum states as well as the
definition of a density operator. Also here, \cite{ACS} has managed to
implicitly choose a quantization that is beneficial in producing such a
bound. Inverse-volume corrections do not change this qualitative picture, as
shown in Sec.~\ref{s:Inv} mainly using a new set of useful analog models
introduced in Sec.~\ref{s:Analog}. An interesting feature of non-bouncing
solutions is that they lead to novel examples of quantum space-time
structures, possibly avoiding signature change.

\section*{Acknowledgements}

This work was supported in part by NSF grant PHY-1607414.


\end{document}